\def\cm{cm$^{-1}$}
\def\bfa{Ba\-Fe$_2$As$_{2}$}

\def\bfca{Ba\-(Fe$_{0.92}$Co$_{0.08})_2$As$_{2}$}
\def\bfna{Ba\-(Fe$_{0.95}$Ni$_{0.05})_2$As$_{2}$}

\documentclass[prl,twocolumn,showpacs,amsmath,amssymb]{revtex4}
\usepackage[dvips]{graphicx}
\usepackage{graphicx}
\usepackage{color}

\begin{document}
\title{Eliashberg Analysis of Optical Spectra Reveals
Strong Coupling of \\
Charge Carriers to Spin Fluctuations in Superconducting
Iron Pnictides }
\author{D. Wu}
\author{N. Bari\v{s}i\'{c}}
\author{M. Dressel}
\affiliation{1.~Physikalisches Institut, Universit\"at Stuttgart,
Pfaffenwaldring 57, 70550 Stuttgart, Germany}
\author{G. H. Cao}
\author{Z-A. Xu}
\affiliation{Department of Physics, Zhejiang University, Hangzhou 310027, People's Republic of China}
\author{E. Schachinger}
\affiliation{Institut f\"ur Theoretische Physik, Technische Universit\"at Graz,
Petersgasse 16, 8010 Graz, Austria}
\author{J. P. Carbotte}
\affiliation{Department of Physics and Astronomy, McMaster University, Hamilton, Ontario L8S 4M1, Canada}
\date{\today}

\begin{abstract}
The temperature and frequency dependences of the optical conductivity of Co and Ni-doped BaFe$_2$As$_2$ are analyzed and
the electron-boson spectral density $\alpha^2F(\omega)$ extracted using Eliashberg's formalism.
The characteristic energy of a large peak in the spectrum around 10~meV
coincide with the resonance peak in the spin excitation spectra, giving
compelling evidence that in iron-based superconductors spin fluctuations
strongly couple to the charge carriers and mediate superconductivity.
In addition the spectrum is found to evolve with temperature towards
a less structured background at higher energies as in the spin susceptibility.
\end{abstract}

\maketitle

Whenever a new family of superconductors is discovered, among the first questions posed are about the mechanism of superconductivity \cite{Kivelson08,Mazin10}. In conventional superconductors,
phonons mediate the attractive interaction of two electrons forming Cooper pairs. For heavy fermions, cuprates and some organic  superconductors, magnetic Cooper-pairing mechanisms have been proposed as a candidate \cite{BennemannKetterson03}. However, intrinsic complications have prevented the general acceptance. The situation of the novel class of iron based superconductors is comparably unsettled \cite{Ishida09}. Nevertheless, the proximity of superconductivity and antiferromagnetism in the phase diagram \cite{Lynn08}, the weak electron-phonon interaction \cite{Boeri08,Choi10} and the resonance peak in the spin-excitation spectrum \cite{Inosov10,Wang10} support the hypothesis of a magnetic interaction leading to superconductivity \cite{Mazin08,Kuroki08,Vorontsov08}.

The Eliashberg spectral function $\alpha^2F(\omega)$ quantifies
the boson exchange effect and is a good way to discriminate
between the candidate mechanisms. For half a century,
current-voltage characteristics obtained by tunneling spectroscopy
are utilized to provide detailed and accurate information on the
phonon exchange for most conventional superconductors
\cite{Carbotte90,Parks69}. In strongly correlated systems, such as
high-$T_c$ cuprates, the inversion of optical data is commonly
used to extract the bosonic excitation spectra
\cite{Carbotte99,Basov05,Hwang08}. The analysis and
interpretation, however, is not straight forward since a certain
complexity in understanding the spectral signature of bosonic
modes arises from the joint mechanisms in these particular
systems. In this regard, Carbotte and collaborators succeeded to
modify the kernel $\alpha^2F(\omega)$ of the Eliashberg theory by
introducing the nearly antiferromagnetic Fermi-liquid model where
the exchanged bosons are described as antiferromagntic spin
fluctuations, and the resulted optical resonance tracked very well
the temperature evolution of the spin resonance
seen in neutron  scattering \cite{Carbotte99}.


In this Letter, we reported a detailed analysis of our optical
spectra obtained on \bfca\ single crystals. The inversion of the
frequency dependent optical scattering rate $\tau^{-1}(\omega)$ reveals
that the coupling of charge carriers to bosonic modes has an
optimum peak around 10~meV, with a coupling constant $\lambda=4.4$
right above $T_c$. With increasing temperature, this peak becomes
broader and moves to higher energies. These bosonic spectral
signatures closely resemble the dynamical spin susceptibility
$\chi^{\prime\prime}(\omega)$ observed by neutron  scattering,
indicating a magnetic mediation mechanism in the novel iron-pnictide
superconductors.

\begin{figure}
 \includegraphics[width=0.9\columnwidth]{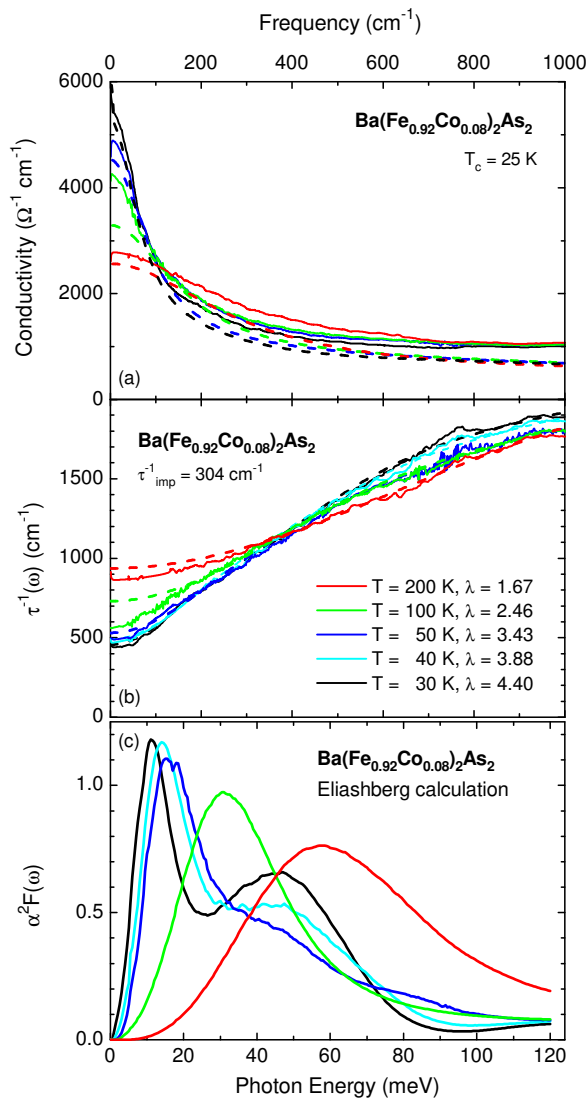}
 \caption{\label{fig:Fig1} (Color online) Results of the inversion calculations of the electron-boson spectral function $\alpha^2F(\omega)$ for \bfca\ at
 different temperatures in the normal state. (a)~Measured conductivity spectra (solid lines) compared with computational results (dashed lines).
(b)~Frequency dependent scattering rate obtained
by the extended Drude analysis of the conductivity spectra plotted in the upper panel (solid lines) compared with the calculated $1/\tau(\omega)$ according to Eq.~(\ref{eq:alpha2F}) (dashed lines) with a constant contribution of 304~\cm\ due to impurities. (c)~The corresponding electron boson spectral density $\alpha^2F(\omega)$ calculated by Eliashberg formalism.}
\end{figure}

We have measured the optical reflectivity of Co-doped \bfa\ single crystals over a wide frequency and temperature range as described in detail in Ref.~\onlinecite{Barisic10}. The samples are well characterized and exhibit a superconducting transition at $T_c=25\,$K \cite{Wu10}. Via Kramers-Kronig analysis we calculate the complex conductivity $\hat{\sigma}=\sigma_1 +{\rm i}\sigma_2$
which is further analyzed by the extended Drude model in order to obtain the frequency dependent
scattering rate $1/\tau(\omega)$ and mass enhancement $m^*(\omega)/m_{b}=1+\lambda(\omega)$ compared to the bandmass $m_b$. The results are plotted in Fig.~\ref{fig:Fig1} for different temperatures.

In principle, optical data are encoded with information
on the microscopic interaction between the charge carriers. For an
electron-boson system, the Eliashberg equations apply and a Kubo formula can be
used to get the infrared conductivity from the electron-boson spectral
density once the quasiparticle self-energy $\Sigma(\omega,T)$ is known
\cite{Lee84}. The opposite direction turns out to be even more challenging
since we have to invert an equation of the form
\begin{equation}
\frac{1}{\tau(\omega)}=\frac{1}{\tau_{\rm imp}}+\int^\infty_0 K(\omega,\Omega;T)\,\alpha^2F(\Omega)\, {\rm d}\Omega  \quad ,
\label{eq:alpha2F}
\end{equation}
where $1/\tau_{\rm imp}$ denotes a constant scattering rate due to impurities; the normal state kernel is given by \cite{Shulga37}:
\begin{eqnarray}
&&K(\omega,\Omega;T)=\frac{\pi}{\omega}\left[ 2\omega \coth\left\{\frac{\hbar\Omega}{2k_BT}\right\}\, - \right.
\\
&&\left.(\omega-\Omega)
\coth \left\{\frac{\omega+\Omega}{2k_bT/\hbar}\right\} +(\omega-\Omega)\coth \left\{\frac{\omega-\Omega}{2k_bT/\hbar}\right\} \right]~.
\nonumber
\end{eqnarray}
Several methods have been suggested to extract the information on the
electron-boson spectral density $\alpha^2F(\omega)$ from the optical
scattering rate, such as singular value decomposition, maximum entropy method
and least square fit; a detailed discussion of advantages and limitations of
these numerical inversion techniques is given in
Ref.~\cite{Schachinger06}. Here we have approached the deconvolution by
an unbiased maximum entropy method, similar to \cite{Hwang08}, and plot the outcome in Fig.~\ref{fig:Fig1}(c) for various temperatures
in the metallic state of \bfca.

Note, when a gap in the density of states opens below $T_c$, the present
analysis of $\tau^{-1}(\omega)$ becomes meaningless; hence we have to restrict
ourselves to $T>T_c$. However, it is safe to assume that electron-boson
coupling makes a strong impact on the spectra already in the normal state just
above $T_c$, and it is this quantity which will determine $T_c$. Before
starting the discussion, we would like to emphasize that the
Eliashberg inversion applied here is based on single band systems with
infinite band width while most of materials actually have
finite band width. The iron pnictides, on the other hand, are
certainly multiband systems (see for instance Refs. \cite{Mazin08} and
\cite{Vorontsov08}). In the normal state the optical conductivity of such
a multi-band system is just the sum of the various band contributions
to this conductivity, provided the interband optical transitions are zero or
sufficiently small, as was suggested by van Heumen {\it et al.}
\cite{Heumen09} for the Ba(Fe$_{1-x}$Co$_x$)$_2$As$_2$ class of materials.
As is generally believed, the inelastic
scattering is dominated by interband transitions due to possibly spin
fluctuations and, thus, a single form of the electron-boson spectral density
will enter the problem except for a possible scaling factor
accounting for a different magnitude of $\alpha^2F(\omega)$ for
transitions between different bands. This was considered explicitly
by Benfatto {\it et al.} \cite{Benfatto09}. Thus, the application
of Eq.~\eqref{eq:alpha2F} for the deconvolution of experimental data
will yield meaningful information about the electron-boson interaction
in such systems and will provide an {\em average} $\alpha^2F(\omega)$
spectrum. The shape of such a spectrum will still provide meaningful
information on the bosons responsible for superconductivity, for
example phonons or spin fluctuations. A knowledge of its average
magnitude is also equally important. Finally, the use of
a formula based on infinite band width to deconvolute optical data
of systems with a rather narrow band width makes it extremely important to
check the Kramer-Kronig consistency of $\alpha^2F(\omega)$ with optical
constants to exclude any incorrect solutions due to finite band-width
effects. In order to demonstrate the applicability of our analysis, we added
the calculated $\sigma_1(\omega)$ and $\tau^{-1}(\omega)$ to
Fig.~\ref{fig:Fig1}(a) and (b) as dashed lines; both cases show good agreement between theory and experiment. This gives confidence in the
physical relevance of the derived spectra.

In Fig.~\ref{fig:Fig1}(c), one sees a clear temperature dependence
to the recovered electron-boson spectra which is also in agreement with a second, biased maximum entropy inversion
\cite{Hwang08}. At
temperatures just above $T_c$, a pronounced peak centered at 10~meV and a
shoulder around 45~meV dominate the spectral weight below 80~meV. When $T$
increases, this peak moves to higher energies and smears out
quickly as a shoulder. As a consequence, the mass renormalization factor
$\lambda$  is reduced from 4.4 to 1.67 at $T=200$~K.
Since 10~meV seems to be a reasonable energy for phonon excitations,
at first glance, it is tempting to consider a phonon mechanism for superconductivity.
However, compared to band structure calculations \cite{Boeri08,Choi10}, the observed spectral features are quite different. Moreover, from calculations of the electron-boson excitation spectra a characteristic phonon frequency $\omega_{ln}$ can be extracted; in our case $\hbar\omega_{ln}=14.2$~meV. When the phonon mechanism is dominant in a superconducting material, one can estimate the coupling strength by its ratio to  $T_c$. Here, we obtain $k_BT_c/\hbar\omega_{ln}=0.15<0.25$, implying a conventional strong coupling material; it also yields $\lambda$ to be in the range  $1-2$, according to the McMillan equation \cite{Carbotte90}. Obviously, this is much too small compared to our experimental result ($\lambda=4.4$). Here, we would like to point out that the mass renormalization factor $\lambda$ is widely reported to be $\lambda$=4-5 by other studies \cite{Johnston10}. Thus, we expect another mechanism to play the key role in mediating superconductivity in these materials.

\begin{figure}
 \includegraphics[width=0.9\columnwidth]{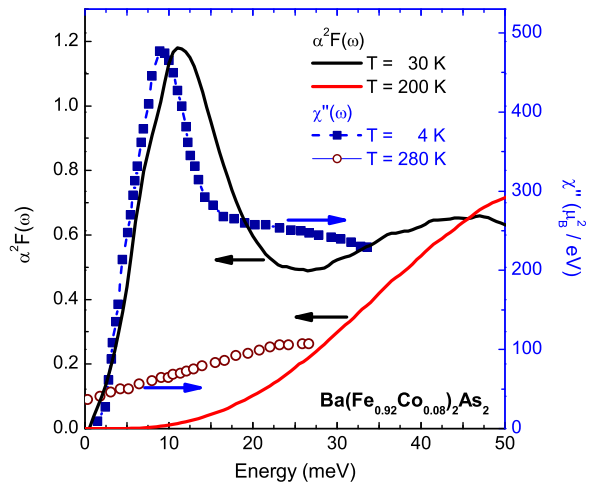}
 \caption{\label{fig:Fig2} (Color online) Comparison of the electron-boson spectral function $\alpha^2F(\omega)$ (solid lines, left axis) derived from inversion of the optical
 scattering rate of \bfca\ and the spin excitation spectrum $\chi^{\prime\prime}(\omega)$ (dots, right axis) obtained from inelastic neutron
 scattering data for ${\bf Q} = {\bf Q}_{\rm AFM}=(\frac{1}{2},\frac{1}{2},1)$ from \protect\cite{Inosov10} at temperatures as indicated.}
\end{figure}

As mentioned above, spin fluctuations seem to be the natural
candidate for the superconducting ``glue'' in iron-based
materials. Whenever a magnetic mechanism is discussed, the main
concern is  whether spin-fluctuation exchange provides sufficient
spectral intensity to make a significant impact on the electronic
self-energy. Very recently, Dahm {\it et al.} \cite{Dahm09}
succeed to establish a quantitative relationship between the
charge- and spin-excitation spectra in high-$T_c$ cuprates, which
demonstrates that the magnetic interaction can generate $d$-wave
superconducting states with transition temperatures comparable to
the maximum $T_c$ observed in these compounds; in other words,
spin fluctuations do have enough strength to cause superconducting
transitions at high-temperature. In Fig.~\ref{fig:Fig2} we display
the electron-boson spectral function $\alpha^2F(\omega)$ derived
from optical  scattering rate of \bfca\ together with the spin
excitation spectrum $\chi^{\prime\prime}(\omega)$ of
Ba\-(Fe$_{0.925}$Co$_{0.075})_2$As$_{2}$ obtained from inelastic
neutron scattering experiment \cite{Inosov10}. Both spectra show a
resonance at approximately 10~meV at low temperature, and the peak
smears out when the temperature increases. Such an important
agreement between optics and neutron indicates that the charge
carriers in this material are strongly coupled  to the spin
excitations.

Our results for \bfna\ ($T_c=20$~K) are qualitatively similar, but
span only a limited energy range and exhibit excessive noise due
to the smaller crystal size \cite{Barisic10,Wu10}. Inelastic
neutron scattering indicates a resonance peak in the spin
excitation spectrum around 7~meV \cite{Wang10} corresponding to
the lower energy scale in this material. Yang {\it et al.}
\cite{Yang09} performed a similar analysis on K doped \bfa\ and
found two maxima of $\alpha^2F(\omega)$ in the range below 30~meV.
Although different in detail, the overall accord gives us
confidence that our observations reveal a general behavior in this
class of materials.

In conclusion, we have analyzed the temperature and frequency dependences of the optical properties of doped BaFe$_2$As$_2$ via Eliashberg theory.
We obtained the electron-boson spectral density $\alpha^2F(\omega)$ which exhibits a characteristic peak around 10~meV.
This coincides with the resonance peak in the spin excitation spectrum and
gives evidence that in iron-based superconductors spin fluctuations strongly
couple to the charge carriers and mediate superconductivity.
Also
there is a strong evolution of the spectra with temperature which agrees
with the known spin fluctuation spectra and this would not be the case
in a phonon mechanism for which the $\alpha^2F(\omega)$ is expected to
remain independent of temperature.

We appreciate discussions with D. N. Basov, A. V. Boris, O. V. Dolgov, B. Gorshunov and D. van der Marel.
D. W. and N. B. acknowledge a fellowship of the Alexander von Humboldt-Foundation.

\end{document}